\documentclass[prl,superscriptaddress,a4paper,twocolumn]{revtex4}
\usepackage{amsmath,amsfonts,graphicx,color}

\begin{document}

\title{Identifying phases of quantum many-body systems that are universal for quantum computation} %
\author{Andrew C. Doherty}%
\affiliation{School of Physical Sciences, The University of Queensland, St Lucia,
  Queensland 4072, Australia}%
\author{Stephen D. Bartlett}%
\affiliation{School of Physics, The University of Sydney,
  Sydney, New South Wales 2006, Australia}%

\date{21 June 2009}

\begin{abstract}
  Quantum computation can proceed solely through single-qubit measurements
  on an appropriate quantum state, such as the ground state of an interacting
  many-body system.  We investigate a simple spin-lattice system based
  on the cluster-state model, and by using nonlocal correlation functions
  that quantify the fidelity of quantum gates performed between distant qubits, we
  demonstrate that it possesses a quantum (zero-temperature) phase transition
  between a disordered phase and an ordered ``cluster phase'' in which it
  is possible to perform a universal set of quantum gates.
\end{abstract}

\maketitle


Measurement-based quantum computation (MBQC) is a fundamentally new
approach to quantum computing.  MBQC proceeds by using only local
adaptive measurements on single qubits.  No entangling operations
are required; all entanglement for the computation is supplied by a
fixed initial resource state  on a
lattice of qubits.  The canonical example of such a resource state
is the so-called \emph{cluster state}~\cite{Rau01,Rau03}. Although a
handful of other universal resources have recently been
identified~\cite{Gro07a,Gro07b,vdN07,Bre08}, there currently exists
very little understanding of precisely which properties of quantum
states allow for universal MBQC.  For example, given a state that is
slightly perturbed from the cluster state, it is not currently known
how to determine if it is a universal resource.  New theoretical
tools are required to identify the properties of potential resource
states that allow for universal MBQC.

A useful perspective to approach this problem is to view the
resource state for MBQC as the ground state of a strongly-coupled
quantum many-body system. With this perspective, we propose that the
ability to perform MBQC is a type of quantum order -- one which can
be identified using appropriate correlation functions as order
parameters.  We show that a natural choice for such correlation
functions are the expectation values of non-local strings of
operators that can be identified with measurement sequences for
performing quantum logic gates within MBQC.  One way of
understanding MBQC is that, by means of a set of local measurements,
it is possible to prepare the resource states required for gate
teleportation~\cite{Got99,Chi05,Rau03} between distant components of
the many-body system.  The performance of the MBQC scheme can be
characterized by calculating the fidelity of the prepared resource
state with the ideal one~\cite{Chu08}.  This fidelity will depend on
a set of non-local correlation functions as a result of the many
local measurements that are required to prepare the resource state.
(Because the fidelity of the identity gate is quantified by the
ability to prepare an entangled state between two distant qubits
using local measurements, it is closely related to the much-studied
property of \emph{localizable entanglement}~\cite{Pop05}.)  We show
that, for the cluster state implementation of MBQC, the specific
correlation functions corresponding to any gate can be calculated,
and we investigate a specific model where the fidelities of a gate
set indeed serve as order parameters identifying a \emph{cluster
phase}.  This result suggests the existence of spin systems that
possess a phase for which any state is a universal resource for
MBQC.  These methods provide new tools for identifying properties of
quantum many-body systems that are required for MBQC.

Consider the following model system.  The cluster state on a lattice
$\mathcal{L}$ is defined as the unique $+1$ eigenstate of a set of
stabilizer operators $K_\mu = X_\mu{\textstyle
\prod_{\nu\sim\mu}}Z_\nu$, where $X_\mu$ ($Z_\mu$) is the Pauli $X$
($Z$) operator at site $\mu$ and where $\nu\sim\mu$ denotes that
$\nu$ is connected to $\mu$ by a bond in the lattice $\mathcal{L}$.
The Hamiltonian $H = -\sum_{\mu \in \mathcal{L}} K_\mu$ has the
cluster state as its unique ground state~\cite{Rau05}. Although the
terms in this Hamiltonian are many-body interactions, it can be
realized as the effective low-energy theory of a Hamiltonian
consisting only of two-body terms~\cite{Bar06}.

As a model system to consider how robust is this Hamiltonian in the
presence of local perturbations, we supplement it with a local field
term,
\begin{equation}
  \label{eq:GeneralClusterHamWithLocalField}
  H(B) = -{\textstyle \sum_{\mu \in \mathcal{L}}} (K_\mu + B X_\mu)\,,
\end{equation}
representing a local transverse field with magnitude $B$.  We refer
to a lattice with this Hamiltonian as the \emph{transverse-field
cluster model} (TFCM), and we will demonstrate the existence of a
single zero-temperature phase transition in the ground state of such
models on both a 1-D line and a 2-D square lattice, separating a
disordered phase from a ``cluster phase''.  Rather than solving
these models explicitly, we explore duality transformations that
relate these models to others with well-understood phases and order
parameters.  We then demonstrate that the order parameters of these
models, mapped back to the TFCM, are precisely equivalent to the
correlation functions in the cluster state that quantify the
fidelity of the identity gate (i.e., teleportation) in MBQC.  That
is, the ability to perform the identity gate over a long range
serves as an order parameter for this phase; similar results hold
for other single-qubit gates as well. In addition, in two
dimensions, we perform a similar analysis of the two-qubit CSIGN
gate, $\exp(i \pi |1\rangle \langle 1|\otimes |1\rangle \langle
1|)$,
which together with our single-qubit gates yields a universal gate
set for MBQC.  (In contrast, the case with a local longitudinal
field instead of a transverse one was investigated in~\cite{LocalZ};
this model demonstrates no such phase but nevertheless can still
allow for MBQC for some range of parameters.)


\textit{General properties of the transverse field cluster model.}
We first present some general properties of the TFCM that are valid
in any dimension and on many lattices, before investigating one- and
two-dimensional models in detail. An immediate observation is that
this model is \emph{self-dual}.  The canonical transformation of
Pauli operators given by applying the CSIGN operation between all
neighbouring pairs of qubits takes $K_\mu \leftrightarrow X_\mu$,
and thus the Hamiltonian (\ref{eq:GeneralClusterHamWithLocalField})
transforms as $H(B) \rightarrow B H(1/B)$.  This self-duality
ensures that, if this model has a single quantum phase transition in
the range $B>0$, then it must occur at $B=1$.

Also, consider lattices which are bipartite, meaning we can divide
the sites into two subsets $\mathcal{L}_{\rm r}$ and
$\mathcal{L}_{\rm b}$, labeled red and blue, such that the
neighbours of any site are all of the other colour. With this
colouring, the Hamiltonian
(\ref{eq:GeneralClusterHamWithLocalField}) can be written as the sum
of two commuting terms, $H = H_{\rm r} + H_{\rm b}$, where
\begin{equation}
  \label{eq:1DClusterHamRed}
  H_{\rm r} = - {\textstyle \sum_{\mu \in \mathcal{L}_{\rm b}}} K_{\mu} - B {\textstyle \sum_{\mu \in \mathcal{L}_{\rm r} }} X_{\mu} \,,
\end{equation}
with $H_{\rm b}$ consisting of the remaining terms.  In the
following, we present mappings of $H_{\rm r}$ (equivalently, $H_{\rm
b}$) in one and two dimensions to known models,  which allows us to
identify the phases and relevant order parameters.

\textit{One dimension.}  Consider the TFCM on a 1-D lattice with fixed
boundary conditions -- a line.  A state of a 1-D lattice cannot serve
as a universal resource for MBQC; however, it will be illustrative to
consider this model as a prelude for studying higher dimensions.  The
Hamiltonian~(\ref{eq:GeneralClusterHamWithLocalField}) on a line with
boundary terms is
\begin{multline}
  \label{eq:1DClusterHam}
  H(B) = -{\textstyle \sum_{i=2}^{N-1}} \left( Z_{i-1} X_{i}Z_{i+1} +
    BX_i \right) \\ -X_1Z_2-BX_1 - Z_{N-1}X_N -BX_N \,.
\end{multline}
The ground state of this Hamiltonian is non-degenerate, and for
$B=0$ is given by the 1-D cluster state on a line.  Pachos and
Plenio~\cite{Pac04} have shown explicitly that this model (with
periodic boundary conditions) exhibits a quantum phase transition at
$|B|=1$, and that the localizable entanglement length remains
infinite for all values $|B|<1$. Their method makes use of the
Jordan-Wigner transformation to yield a linear fermionic system.  We
provide a more direct transformation to a known model -- the
transverse-field Ising model~\cite{Kog79} -- which provides a
natural generalization to higher-dimensional lattices.

\begin{figure}
  \includegraphics[width=3in]{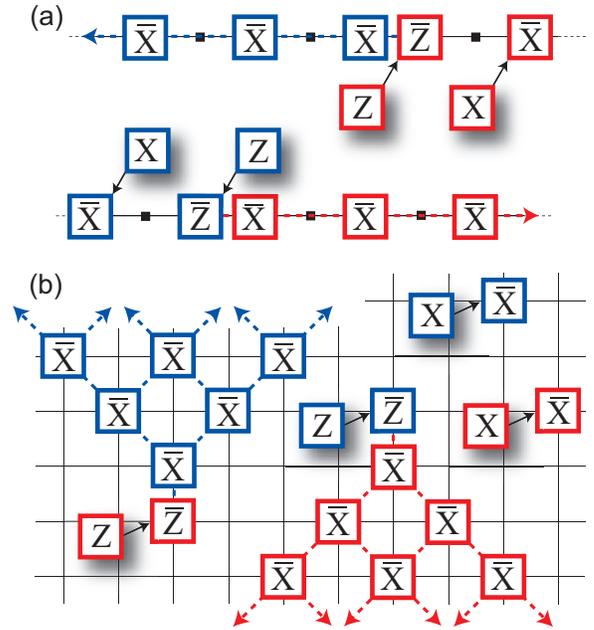}
  \caption{(a) The duality transformation of
  Eqs.~(\ref{eq:1Dred}-\ref{eq:1Dblue}) on a 1-D line.  (b) A
  generalization of this duality transformation to a 2-D square
  lattice.}
  \label{fig:DualityTrans}
\end{figure}

Our duality transformation is as follows.  On red (even) sites, the
Pauli operators transform as
\begin{equation}
  \label{eq:1Dred}
  X_{2j}\to \bar{X}_{2j}\,, \quad Z_{2j} \to \bigl({\textstyle\prod_{k=1}^{j}} \bar{X}_{2k-1} \bigr)\bar{Z}_{2j}\,.
\end{equation}
On blue (odd) sites, the Pauli operators transform as
\begin{equation}
  \label{eq:1Dblue}
  X_{2j-1}\to \bar{X}_{2j-1}\,, \quad Z_{2j-1} \to \bar{Z}_{2j-1}\bigl({\textstyle\prod_{k=j}^{N}} \bar{X}_{2k} \bigr)\,.
\end{equation}
This mapping is canonical, meaning the new Pauli matrices
$\bar{X}_j$ and $\bar{Z}_j$ satisfy the correct commutation and
anti-commutation relations.  An illustration of this transformation
is presented in Fig.~\ref{fig:DualityTrans}(a).  We emphasize that
this duality transformation is non-local, and thus the properties of
a system for MBQC are not preserved under this mapping.  However, as
we now demonstrate, the phases and order parameters of this dual
model are well-studied and will allow us to completely classify the
phases as well as calculate the fidelities of the MBQC quantum gates
in the original TFCM.

We consider only the case where $N$ is even.   In terms of
transformed Pauli operators, the Hamiltonian $H_{\rm r}$ acts only
on red sites and has the form
\begin{equation}
  H_{\rm r} = -\bar{Z}_2 - {\textstyle \sum_{i=2}^{N/2}} \bigl( \bar{Z}_{2(i-1)} \bar{Z}_{2i} + B
  \bar{X}_{2i} \bigr) \,,
\end{equation}
The Hamiltonian $H_{\rm b}$ is similar, acting only on blue sites,
with a $\bar{Z}$ boundary term at $j=N$. This mapping on the TFCM,
then, yields two identical transverse-field Ising models, one on
each of $\mathcal{L}_{\rm r}$ and $\mathcal{L}_{\rm b}$. Each has a
local $\bar{Z}$ field term which breaks the symmetry in the ordered
($|B|<1$) phase and specifies a unique ground state.  The ground
state of the total lattice is then non-degenerate and is given by
the product state of these two unique ground states.

The solution to this known model allows us, via the duality
transformation, to completely characterise the TFCM. For example,
the phases of the TFCM are specified by the well-studied phases of
the transverse-field Ising model; in particular, there is a unique
quantum phase transition at $|B|=1$~\cite{Pfe70}.  Also, the
well-known order parameters for the transverse-field Ising model can
be mapped, using the duality transformation, to order parameters for
the TFCM.  In the ordered phase of the transverse-field Ising model,
the correlation functions $\langle \bar{Z} \bar{Z}\rangle$ (for both
colors) are long ranged. (Specifically, $\lim_{k\to\infty} \langle
\bar{Z}_i \bar{Z}_{i+k}\rangle = (1-|B|^2)^{1/4}$ for
$|B|<1$~\cite{Pfe70}.) We can use this result to make a
corresponding statement about correlation functions for the TFCM. By
reversing the duality transformation, we have
\begin{align}
  \label{eq:IsingOrderBlue}
  \langle \bar{Z}_{2i-1} \bar{Z}_{2j-1} \rangle
  &\rightarrow \langle Z_{2i-1}
  ({\textstyle\prod_{k=i}^{j-1}} X_{2k} )
  Z_{2j-1} \rangle = \langle {\textstyle\prod_{k=i}^{j-1}} K_{2k} \rangle \,, \\
  \label{eq:IsingOrderRed}
  \langle \bar{Z}_{2i} \bar{Z}_{2j} \rangle
  &\rightarrow \langle Z_{2i}
  ({\textstyle\prod_{k=i}^{j-1}} X_{2k+1} )
  Z_{2j} \rangle = \langle
  {\textstyle\prod_{k=i}^{j-1}} K_{2k+1}
  \rangle\,.
\end{align}
That is, in the phase $|B|<1$ wherein $\langle
\bar{Z}\bar{Z}\rangle$ is long-ranged, the string-like operators
corresponding to the product of even (or odd) stabilizers $K_i$ in
the TFCM are also long-ranged, with the limiting value
$(1-|B|^2)^{1/4}$.  These two correlation functions are all that is
needed to calculate the fidelity of the resource state for the
identity gate with the ideal maximally-entangled state, and it is
found to be $>1/4$ for all $|B|<1$.  (The average fidelity of a
randomly chosen state yields $1/4$.)  The same calculation for other
single-qubit Clifford gates~\cite{nielsen2000a} and for an arbitrary
$Z$-rotation $U_z(\theta) = \exp(-i\theta Z)$ (a non-Clifford gate),
yields the same result~\cite{Chu08}.

Thus, this duality transformation has allowed us to prove our
desired results:  First, that the TFCM does indeed possess a phase,
given by $|B|<1$, which we denote the \emph{cluster phase}.  The
order parameters of this phase, given by products of even or odd
stabilizer operators $K_i$, demonstrate that quantum gates can be
performed with high fidelity (relative to a randomly-chosen state)
using any state within this phase.  The ground states in this phase
are indeed ``robust'' against variations in the precise value of
$B$.  However, the one-dimensional cluster state is not a universal
resource for MBQC, and so we direct our attention to a
two-dimensional model.

\textit{Two dimensions.}  We consider a square lattice; the cluster
state on this lattice is a universal resource for MBQC.  This
lattice is bipartite, and thus we can define the commuting
Hamiltonians $H_{\rm r}$ and $H_{\rm b}$ as above.  We use a natural
generalization of the 1-D duality transformation, as follows.  On
red sites, Pauli operators transform as
  $X_{\mu}\to \bar{X}_{\mu}$ and $Z_{\mu} \to \bigl({\textstyle\prod_{\mu'>\mu}} \bar{X}_{\mu'} \bigr)\bar{Z}_{\mu}$,
whereas on blue sites,
  $X_{\mu}\to \bar{X}_{\mu}$ and $Z_{\mu} \to
\bar{Z}_{\mu}\bigl({\textstyle\prod_{\mu'<\mu}} \bar{X}_{\mu'}
\bigr)$.
Here, $\mu'>\mu$ ($\mu'<\mu$) denotes that $\mu'$ lies in the upper
(lower) cone relative to $\mu$ as in Fig.~\ref{fig:DualityTrans}(b).
Again, one can easily verify that this transformation is canonical.

Under this mapping, each stabilizer maps to a monochromatic operator
consisting only of $\bar{Z}$ terms.  Non-boundary stabilizers map to
products of four $\bar{Z}$ operators on the corners of a fundamental
plaquette $\Box$; boundary conditions can be chosen such that
boundary stabilizers map to two-$\bar{Z}$ and one-$\bar{Z}$ terms.
The Hamiltonian $H_{\rm r}$ ($H_{\rm b}$) on $\mathcal{L}_{\rm r}$
($\mathcal{L}_{\rm b}$) maps to
\begin{equation}
    \label{eq:Ham2DJWnoboundary}
  H = - \sum_{\Box} \begin{matrix} \bar{Z} & \bar{Z} \\ \bar{Z} & \bar{Z} \end{matrix}
  - B \sum_{\mu} \bar{X}_\mu \,,
\end{equation}
plus boundary terms (not shown) which ensure a non-degenerate ground
state for all $B$.  This model possesses a phase transition at
$|B|=1$~\cite{xu2004a,xu2004b}. Thus, through this duality map, we
know that the 2-D TFCM has a phase transition at $|B|=1$, and we use
the term \emph{cluster phase} to denote the $|B|<1$ phase.  In
addition, this model of Eq.~(\ref{eq:Ham2DJWnoboundary}) is dual to
the \emph{anisotropic quantum orbital compass model}
(AQOCM)~\cite{nussinov2005a,doucot2005a,dorier2005a,bacon2005a},
with a mapping that also locally maps the boundary terms).  The key
advantage of the AQOCM is that it contains only two-body terms in
the Hamiltonian, and is therefore very amenable to numerical
investigation.  For example, the projected entangled-pair state
algorithm applied to this model provides very strong evidence that
the phase transition is first order~\cite{orus2009a}. The model also
possesses correlation functions for an Ising order parameter that
simulations indicate are long-ranged for $|B|<1$~\cite{orus2009a}.

\begin{figure}
 \includegraphics[width=3.25in]{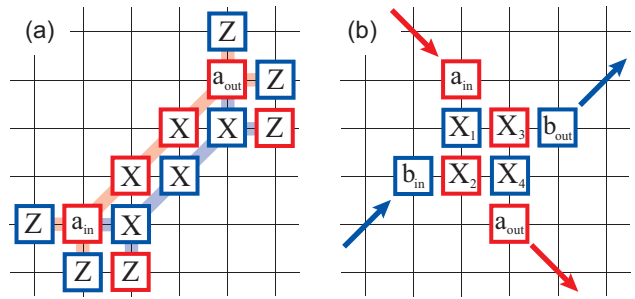}
\caption{(a) A measurement pattern on the cluster state that
localizes entanglement between sites ${\rm a}_{\rm in}$ and ${\rm
a}_{\rm out}$, where $X$ ($Z$) denotes a measurement in the
$X$-basis ($Z$-basis). The two string-like stabilizers, centred on
sites connected by the shaded red and shaded blue diagonal lines,
have long-ranged expectation values in the $|B|<1$ phase; these
correlation functions directly quantify the fidelities of
single-qubit gates between ${\rm a}_{\rm in}$ and ${\rm a}_{\rm
out}$ in MBQC.   (b)  The measurement sequence corresponding to the
CSIGN gate between ${\rm a}$ and ${\rm b}$. The expectation of four
stabilizers characterises the CSIGN gate: $K_{{\rm a}_{\rm
in}}K_3K_{{\rm a}_{\rm out}}$, $K_{{\rm b}_{\rm in}}K_4 K_{{\rm
b}_{\rm out}}$, $K_1 K_4$ and $K_2 K_3$. These stabilizers can be
appended with diagonal strings of red (blue) stabilizers in the
direction of the arrows (and terminated with $Z$ measurements as in
(a)) to reach distant qubits.  With $X$ measurements on qubits 1-4,
the resulting state provides the CSIGN transformation.}
 \label{fig:2Dgates}
\end{figure}

Inverting this duality transformation and returning to the TFCM,
these Ising-type correlation functions map onto strings of
monochromatic stabilizers along diagonal lines in the square lattice
(see Fig.~\ref{fig:2Dgates}(a)).  Again using the correlation
functions for single-qubit gates given in~\cite{Chu08}, we find that
these strings of monochromatic stabilizers characterize the
fidelities of the identity gate and a generating set of single-qubit
gates between two distant points, and serve as order parameters for
the cluster phase.

In addition, in this 2-D model we can consider two-qubit gates.  We
make use of the elementary measurement pattern for a CSIGN gate on
two qubits which are subsequently swapped, as given in
Refs.~\cite{Rau03,Chu08} and shown in Fig.~\ref{fig:2Dgates}(b). The
desired long-ranged correlation functions on the AQOCM are of the
form of 4-body correlators $\langle \tilde{Z}_{(i,j_0)}
\tilde{Z}_{(i,j_*)} \tilde{Z}_{(i+1,j_*)} \tilde{Z}_{(i+1,j_1)}
\rangle$, where $j_*$ is an intermediate column between $j_0$ and
$j_1$.  Such 4-body correlation functions should be possible to
numerically evaluate in the AQOCM using recent techniques.  The
CSIGN together with the above single-qubit gates yields a universal
gate set, and thus the cluster phase is indeed characterized by the
fidelities of a universal gate set for MBQC.

\textit{Discussion.}  Using the TFCM as an example, we have
demonstrated the utility of correlation functions corresponding to
quantum gates as order parameters to identify a phase according to
its usefulness for MBQC.
The perspective of quantum-computational universality of a
state as a new type of quantum order may assist in identifying new
quantum systems that can be used for MBQC.

The behaviour of the TFCM contrasts with the model considered
in~\cite{LocalZ}, which is the cluster state Hamiltonian perturbed
by a local $Z$-field. In that model the gate correlation functions
discussed in this paper become short-ranged at any non-zero
perturbation. However, by pre-processing with certain local
filtering operations it is still possible to perform MBQC for
sufficiently low field and sufficiently low
temperature~\cite{LocalZ}. Unlike the TFCM, the model
of~\cite{LocalZ} does not undergo a phase transition. These
behaviours are very reminiscent of the quantum Ising model in one
dimension; where there is a broken symmetry that disappears at a
phase transition for sufficiently large transverse field but
longitudinal fields destroy the ground state order without any phase
transition.

One could also ask whether these ordered phases persist to finite
temperature.  As our model is gapped except at the phase transition,
it is possible for a finite-sized thermal system to be cooled to
have arbitrarily high overlap with the ground state (although this
becomes a challenge close to the phase transition).  In one
dimension, the fact that the transverse-field Ising model does not
maintain an ordered phase at any finite temperature demonstrates
that the 1-D TFCM does not either.  In two dimensions, it is less
clear.  For this reason it would be worth investigating the TFCM on
a three-dimensional lattice such as in~\cite{Rau05}, for which the
$B=0$ model is known to allow for fault-tolerant MBQC at finite
temperature~\cite{Rau05,LocalZ,Rau06}.

\begin{acknowledgments}
  \textit{Acknowledgments.}  We acknowledge helpful discussions with
  Sean Barrett, Dan Browne and Terry Rudolph, and the support of the
  Australian Research Council.
\end{acknowledgments}

\end{document}